\documentclass[12pt]{article}
\usepackage{titling}
\usepackage{authblk}
\usepackage{setspace}
\usepackage{ragged2e}
\usepackage{fancyhdr}
\usepackage{multicol}
\usepackage{graphicx}
\usepackage{caption}
\usepackage{amsmath} 
\usepackage{fancybox} 
\usepackage{array}
\usepackage{multirow}
\usepackage{subcaption} 
\usepackage{placeins}
\usepackage{float}
\usepackage{url}
\usepackage{hyperref}
\usepackage{natbib}
\usepackage{hyperref}
\hypersetup{
	colorlinks=true,           
	linkcolor=blue,            
	citecolor=blue,           
	filecolor=magenta,         
	urlcolor=magenta               
}

\title{Robust Feature Engineering Techniques for Designing Efficient Motor Imagery-Based BCI-Systems}
\justifying
\author[1]{\small Syed Saim Gardezi}
\author[2]{\small Soyiba Jawed\thanks{Corresponding author: Soyiba Jawed (email: soyiba.jawed@ceme.nust.edu.pk)}}
\author[3]{\small Mahnoor Khan}
\author[4]{\small Muneeba Bukhari}
\author[5]{\small Rizwan Ahmed Khan}
\justifying
\affil[1]{\footnotesize Department of Biomedical Engineering, Faculty of Engineering, Salim Habib University, Karachi, Pakistan}
\affil[2]{\footnotesize Department of Computer Software Engineering, College of Electrical and Mechanical Engineering, NUST, Islamabad, Pakistan}
\affil[3]{\footnotesize Department of Physics, Faculty of Applied Sciences, Riphah International University, Islamabad, Pakistan}
\affil[4]{\footnotesize Department of Biomedical Engineering, Faculty of Engineering, NED University, Karachi, Pakistan}
\affil[5]{\footnotesize Department of Computer Science, School of Mathematics and Computer Science, Institute of Business Administration, Karachi, Pakistan}
\date{}

\begin{document}
\maketitle
\vspace{-1.5cm}
\begin{abstract}
\justifying
A multitude of individuals across the globe grapple with motor disabilities. Neural prosthetics utilizing Brain-Computer Interface (BCI) technology exhibit promise for improving motor rehabilitation outcomes. The intricate nature of EEG data poses a significant hurdle for current BCI systems. Recently, a qualitative repository of EEG signals tied to both upper and lower limb execution of motor and motor imagery tasks has been unveiled. Despite this, the productivity of the Machine Learning (ML) Models that were trained on this dataset was alarmingly deficient, and the evaluation framework seemed insufficient. To enhance outcomes, robust feature engineering (signal processing) methodologies are implemented. A collection of time domain, frequency domain, and wavelet-derived features was obtained from 16-channel EEG signals, and the Maximum Relevance Minimum Redundancy (MRMR) approach was employed to identify the four most significant features. For classification K Nearest Neighbors, Support Vector Machine, Decision Tree, and Naïve Bayes models were implemented with these selected features, evaluating their effectiveness through metrics such as testing accuracy, precision, recall, and F1 Score. By leveraging SVM with a Gaussian Kernel,  a remarkable maximum testing accuracy of 92.50\% for motor activities and 95.48\% for imagery activities is achieved. These results are notably more dependable and gratifying compared to the previous study, where the peak accuracy was recorded at 74.36\%. This research work provides an in-depth analysis of the MI Limb EEG dataset and it will help in designing and developing simple, cost-effective and reliable BCI systems for neuro-rehabilitation.

\end{abstract}

\justifying
\section{Introduction}
According to the World Health Organization, globally 1.3 billion people (16\% of total world’s total population) have significant disabilities 
 \cite{world2022global}. Disabled people face difficulty in accessing public transport, hesitate to attend public gatherings, and are more sensitive to different diseases such as obesity, depression, diabetes asthma etc. \cite{sabrin2022biopsychosocial}. Rehabilitation engineering is the domain of Biomedical Engineering in which theoretical knowledge is applied in designing, developing, adapting, testing, evaluating, and distributing technological solutions to problems confronted by individuals with disabilities \cite{cooper2006introduction}.  Because of the advancements in signal processing and intelligence, researchers are rapidly developing brain-computer-based neural prostheses to replace or restore lost nervous functions. 

A brain-computer interface (BCI) system monitors and decodes neurophysiological signals and produces computer commands to control either a single output device or a variety of other devices. More simply, a BCI can be defined as “a system that translates brain signals into new kinds of outputs” \cite{wolpaw2012brain}. These devices help to perform a range of activities including neuroplasticity and controlling assistive devices. The categorization of BCI systems is most often based on whether they employ invasive or non-invasive approaches in the measurement of brain activities \cite{khan2020review}. The definition of invasive BCI systems can be characterized by either direct placement of electrode arrays on the brain surface for electrocorticography (ECoG) or the insertion of micro-electrode arrays into the brain cortex. Researchers working with brain surface electrodes in BCI systems have used epidural electrodes (that are placed between the spinal cord and the vertebrae) \cite{murguialday2011transition}, and subdural electrodes (that are inserted through a small opening in the skull and implanted into a specific brain area) \cite{schalk2007decoding}, \cite{schalk2008two}.  Because of the problems related to the long-term robustness of acquired signals and several ethical and legal concerns, invasive BCI systems yield limited success in real biological settings \cite{suner2005reliability}. Electroencephalography (EEG)-based non-invasive systems are lightweight, safer, more comfortable, and more economical, and hence they are becoming the popular option for acquiring relevant brain signals. In such systems, numerous electrodes are placed on the scalp to acquire the EEG signals. From these signals, features regarding the user’s movement intention are extracted and used to control specific motor devices \cite{grozea2011bristle}.

In order to interface computers with the brain, brain activities are analyzed by acquiring the EEG signals. EEG-based BCI systems are categorized into four main types, depending on how the brain signals are extracted \cite{nicolas2012brain}. These types include Steady State Visual Evoked Potential (SSVEP), P300, Slow Cortical Potential (SCP), and Motor Imagery. Users produce SSVEP by inputting visual stimulation of the flashlights at precise frequencies. The EEG system captures the brain activities that are aligned with these frequencies \cite{wang2008brain}. However, a P300 event-related potential (ERP) is measured from the parietal lobe and reflects brain responses about 300 milliseconds (ms) after receiving visual, auditory or somatosensory stimuli \cite{donchin2000mental}. Slow Cortical Potentials are the sustained changes in membrane potentials of the cortex, which can last from one to several seconds. Positive SCPs show decreased neuronal activity, while negative SCPs indicate increased neuronal activity \cite{strehl2009slow}. It’s significant to highlight that SSVEPs respond to the entire stimulation duration, whereas ERPs are a reaction to a specific event of stimuli. Motor Imagery (MI) based BCI systems are particularly useful for stroke patients whose limbs may not respond adequately to the stimuli produced by the brain. MI consists of visuomotor imagination, requiring users to imagine motor tasks involving hand or foot movements.  MI is stimulus-independent because here neurological signals drive control behavior without any influence from outside stimuli \cite{neuper2009motor}. 
  
While designing Motor Imagery-based BCI Systems, different types of movements or tasks based on the features extracted from the pre-processed EEG signals are classified through Artificial Intelligence (AI). AI  is the science of making intelligent machines which can perform tasks that require intelligence when performed by humans \cite{ertel2024introduction}. Machine Learning (ML) a domain of AI presents a robust approach to identifying brain patterns related to a specific activity without relying on traditional statistical methods.  ML algorithms can learn from observable data and can automatically identify information that can be used for executing the intended activities (Figure 1). Researchers have utilized publicly accessible datasets for the training of ML algorithms. Because of the small number of subjects, AI models that are trained on these datasets tend to overfit.  Although many more MI-based EEG datasets are available \cite{kaya2018large}, \cite{schirrmeister2017deep}, \cite{schalk2004bci2000} the use of more than 20 electrodes to acquire signals, causes complexity. Additionally, the hardware used is not widely available and it restricts the real-world implementation of the BCI systems for Neurorehabilitation.

\begin{figure}[h]
    \centering
    \includegraphics[width=0.8\textwidth]{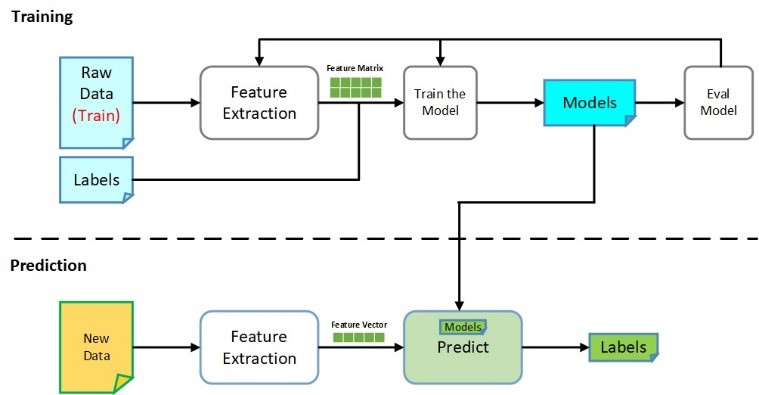} 
    \captionsetup{font=scriptsize} 
    \caption{General Overview of Machine Learning}
\end{figure}

The recent work of  \cite{asanza2023milimbeeg} resolved the above-mentioned challenges by providing a (MI Limb EEG Dataset) related to the execution of motor and motor imagery tasks of the upper and lower limb. This dataset allows us to investigate brain activity patterns during specific motor and motor imagery tasks. Using this dataset, it becomes feasible to train less complex and more robust AI models, mainly for the classification of MI-based movements, which can be valuable in developing and refining rehabilitation techniques and assistive technologies. However, the accuracy of the trained models in previous research was too low and the in-depth analysis of the ML models was not given. To address these challenges, qualitative research was conducted. MI Limb EEG Dataset was used and after digital signal processing, the time domain, frequency domain, and time-frequency domain features were extracted. Feature reduction techniques were applied to reduce the dimensionality and several machine learning algorithms were used for training purposes. The performance of the model was evaluated through metrics such as training accuracy, testing accuracy, precision, recall, and F1 score and was compared with the existing literature for a better understanding of the applications of AI models in the classification of MI-based EEG signals.

The rest of the article is organized as follows: Section 1 presents the literature review. All the details related to dataset, signal processing, feature engineering and classification process are mentioned in Section 2. Section 3 presents the performance of different classifiers. In Section 4 results were discussed and compared with the existing literature, this is followed by the conclusion.

\justifying
\section{Literature Review}
\justifying
MI-based BCI systems face significant challenges due to the complexity and high dimensionality of EEG data. These systems commonly struggle with individual differences in EEG signals, low signal-to-noise ratios, and the non-stationary nature of the data \cite{khademi2023review}.  \cite{gwon2023review} performed a meta-analysis on 25 publicly available EEG Datasets and reported that the mean classification accuracy of Motor Imagery-based BCI systems is only 66.53\%, which turns out to be relatively low. They concluded that a new and qualitative EEG dataset should be made publicly available for designing better BCI systems.

\cite{Umcu2014TitleCT}  suggested that by improving bit rates, improving signal processing techniques and exploring new classiﬁcation approaches better BCI systems could be developed.  Vansteensel et al. \cite{vansteensel2017brain}  surveyed by sending a questionnaire to 3500+ BCI researchers worldwide. The focus of 95\% of participants was on EEG-related BCI systems. It was concluded that BCI systems would be commercialized within the next 5-10 years, yet major technological advancements are needed in sensors, overall system performance and user-friendliness.

 \cite{padfield2019eeg}  emphasized that BCI systems must be user-friendly, cost-effective and reliable for success in the commercial market. These systems should be portable, new users should be trained on them in minimum time and most importantly their results should be stable and dependable. Furthermore, they also reported that it is difficult to make good comparisons among the literature because results are not reported in common and standard metrics. BCI systems should be tested on unhealthy/disabled subjects to avoid unrealistic positive results.

Researchers \cite{rodriguez2012automatic}, \cite{ALSAEGH2021102172}, \cite{Kumar} have used Adaptive Autoregressive Modelling (AAM), Power Spectral Density (PSD), and Common Spatial Pattern (CSP) for feature extraction along with Least-Angle regression (LARS) and Principal Component Analysis (PCA) algorithm for feature selection. More focus was placed on supervised machine learning, where labels about MI activity were provided for the training of ML models. Linear discriminant analysis (LDA) and Support Vector Machine (SVM) were the best-performing models but the testing accuracy was below 80\% and, models were also not evaluated through standard metrics. 

Numerous studies \cite{ALSAEGH2021102172}, \cite{alzahab2021hybrid}, are based on small and old datasets for training and validating AI models. BCI competitions are organized regularly where data is provided to validate signal processing and classification methods. Datasets 2a and 2b presented in BCI Competition IV \cite{tangermann2012review} are widely used for MI-based classification. Data set 2a contains EEG signals from nine participants that were acquired using 22 electrodes while the participants were performing MI-based tasks of the left hand, right hand, both feet and tongue, whereas data set 2b contains EEG data collected from nine subjects using 3 bipolar channels during MI based left hand and right-hand activity.  However, due to limited dataset sizes, over-fitting is often reported, where the model performs well on training data but poorly on unseen or new data which results in miss-classification during real-time BCI applications (lack of generalization).

 \cite{asanza2023milimbeeg}  provided a more reliable, less complex and diverse dataset for the classification of motor imagery activity by following proper experimental protocol to avoid degrading the quality. This dataset has the potential to address the above-mentioned challenges. However, the main problem associated with their research work is low accuracy results and lack of in-depth analysis of the performance of AI models. Based on the literature review, we conducted research work to fill this gap. The main goal was to improve the accuracy of ML models for classifying Motor Imagery activities through efficient feature engineering and better optimization techniques along with in-depth and strict evaluation so that better BCI systems could be developed.
\justifying
\section{Methodology}
\subsection{Dataset Description}
\justifying
The MILimbEEG Dataset \cite{asanza2023milimbeeg} used in this study is comprised of EEG signals in the microvolt \(uV\) range and was recorded while performing motor and motor imagery tasks of both upper and lower limbs. Sixty subjects were recruited from Escuela Superior Politécnica del Litoral (ESPOL), research fellows and patients at Guayaquil Luis Vernaza Hospital to help answer the research questions. Among these subjects, one had upper limb amputation, one had lower limb amputation, and one subject had mild intrahepatic cholestasis of pregnancy (ICP). The subjects performed the following motor and motor imagery tasks: Closing Left Hand (CLH), Closing Right Hand (CRH), Dorsal Flexion of Left Foot (DLF), Plantar Flexion of Left Foot (PLF), Dorsal Flexion of Right Foot (DRF), Plantar Flexion of Right Foot (PRF), and Resting in between tasks (Rest) as shown in Figure 2. 

\begin{figure}[h]
    \centering
    \includegraphics[width=0.6\textwidth]{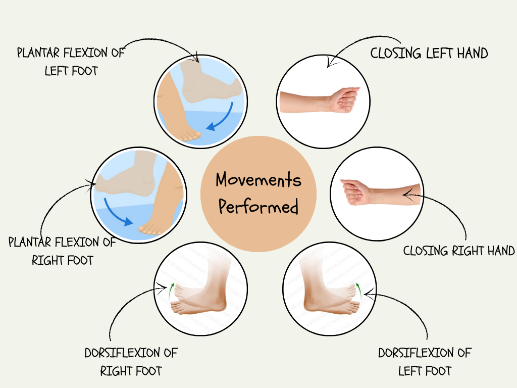} 
    \captionsetup{font=scriptsize} 
    \caption{Motor/ Motor Imagery Movements performed by the subjects}
\end{figure}
The OpenBCI Cyton + Daisy board, in a mono-polar configuration with a sampling rate of 125 Hz, was used to capture the data. The Cyton board has a notch filter at 60 Hz and an active band-pass filter for 5 to 50 Hz \cite{mansouri2023telemetric}. To standardize the recording of the EEG signal and ensure reproducibility, the international 10/10 system standardized by the American EEG Society (AES) was used \cite{seeck2017standardized}. The distribution of the 16 electrodes of the system is shown in Figure 3.

\begin{figure}[h]
    \centering
    \includegraphics[width=0.8\textwidth]{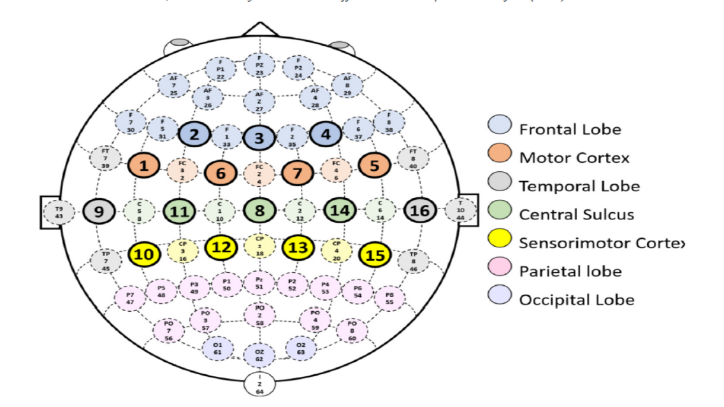} 
    \captionsetup{font=scriptsize} 
    \caption{Distribution of Electrodes while recording the data \cite{asanza2023milimbeeg}}
    \label{fig:your-figure}
\end{figure}

The STL files of the headset used for EEG electrode placement are available at \href{https://github.com/Human-Machine-Interface/OpenBCI_Data_Acquisition}.Data collected on this board were transmitted wirelessly to the OpenBCI software installed on a computer. A Python program via LSL received the EEG data and stored it in comma-separated values (CSV) format \cite{razavi2022opensync}. The data used in the present study includes 2976 CSV files from 24 test subjects. The header row lists the electrodes from 0-15, while each of the remaining 500 rows corresponds to the 500 data points recorded during the task, based on the sample rate. Moreover, each of these files contains 17 columns, with one column corresponding to one sample and the other 16 columns corresponding to the 16 surface electrodes of EEG. Figure 4 shows the general methodology followed during this research work.

\begin{figure}[h]
    \centering
    \includegraphics[width=0.9\textwidth]{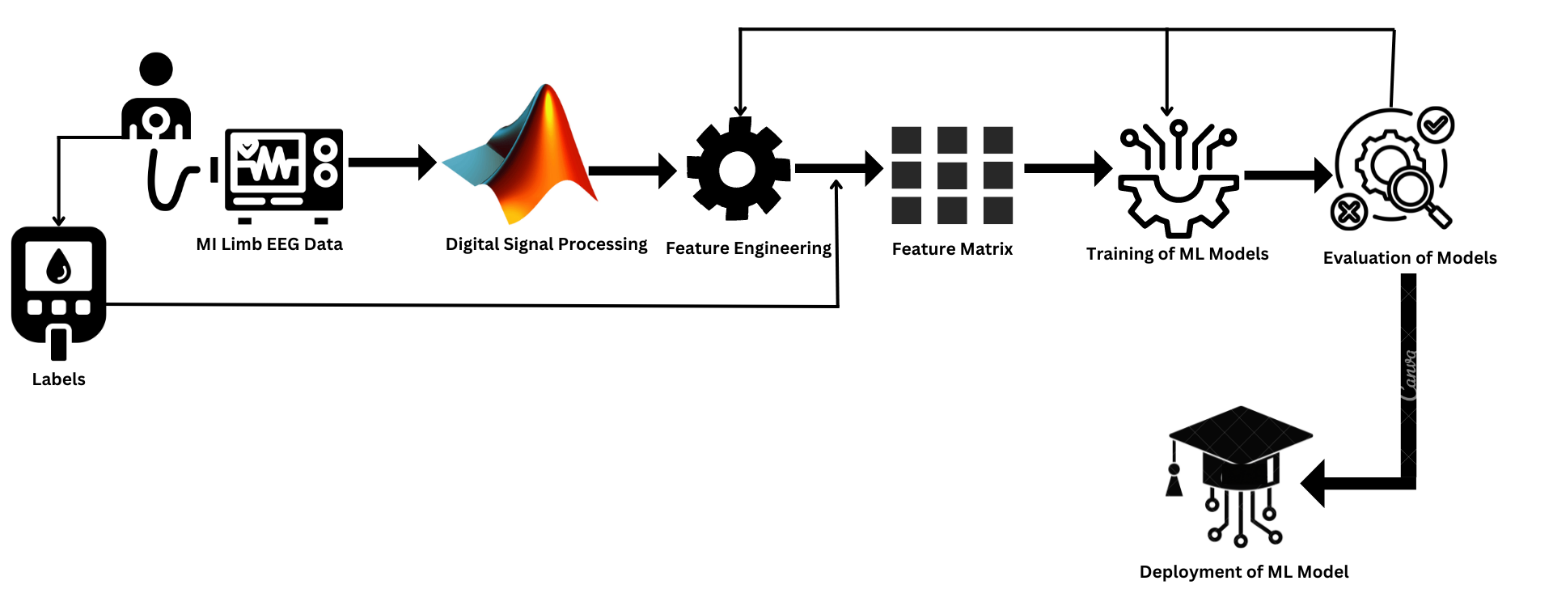} 
    \captionsetup{font=scriptsize} 
    \caption{General Overview of Methodology.}
\end{figure}
\subsection{Preprocessing and Feature Engineering}
\subsubsection{Feature Extraction}
Out of 2,976 CSV files, 16x24=384 files were selected as the main objective was to observe the performance of ML models on comparatively smaller data sizes. Signals in both the time domain and frequency domain were first analyzed, there was no need to apply digital filters at this stage. Normalization was performed by dividing the EEG signal of each channel by its maximum value to ensure that all features have the same scale. Normalization minimizes data variability and thus reduces the chances of over-fitting during the training of the machine learning model. After pre-processing, ten-time domain features were extracted. A brief description of the extracted time domain features is provided in Table 1.

\begin{table}[hp]
    \centering
    \caption{Description of Time Domain Features Extracted from EEG signals}
    \begin{tabular}{|p{3cm}|p{12cm}|}
        \hline
        \textbf{Feature Name} & \textbf{Short Summary} \\
        \hline
        Mean & Mean Amplitude Value of the EEG signal. Raj et al. (2020) identified mean as a feature capable of distinguishing attention-demanding tasks from resting states \cite{raj2020feature}.\\
        \hline
        Standard Deviation (SD) & SD is the square root of variance. It is an effective parameter for comparing real and imagery motion patterns. \\
        \hline
        Mean Energy (ME) & The power of a signal to identify a series in the power curve at any given time is represented by its energy. It provides physiologically rational and empirically effective information about the EEG signals. \\
        \hline
        Mean Teager’s Energy (MTE) & MTE calculates the difference between the squared of the first derivative and the product of the signal and its second derivative. It is used to measure the variation between successive samples of digital signals, and thus it helps to estimate a signal's instantaneous energy. \\
        \hline
        Normalized First Difference (NFD) & The first difference in the normalized EEG signal is the correlation between the current and prior normalized EEG signal data. The NFD helps us to see how quickly or slowly brain activity is changing. \\
        \hline
        Normalized Second Difference (NSD) & This is a measure of how the first difference (the change in signal from one point to the next) changes over time. It is used for analyzing how the speed of brain activity changes. \\
        \hline
        Shannon Entropy (ShEn) & A measurement of the uncertainty (or variability) connected to random variables is called Shannon entropy, or simply entropy. It helps in analyzing the complexity and variability of brain activity. \\
        \hline
        Hjorth Activity (HA) & Hjorth Activity is a time-domain parameter that measures the mean power or activity of a signal. A higher activity value shows more brain activity, while a lower activity value indicates less. \\
        \hline
        Hjorth Complexity (HC) & It gives an estimate of a signal's bandwidth and reveals how closely a signal resembles a pure sinusoidal waveform. This measures the signal's complexity in terms of how many different frequencies are present. A higher complexity value suggests a more intricate signal with diverse frequency components, while a lower complexity indicates a simpler signal. \\
        \hline
        Hjorth Mobility (HM) & The percentage of SD or power spectrum's mean frequency is the Hjorth Mobility. It can be expressed as the square root of the signal's first derivative's activity divided by the signal's activity. Higher mobility means the signal is changing quickly, while lower mobility indicates more stability. \\
        \hline
    \end{tabular}
   
\end{table}

The basic parameters of EEG analysis include the power spectral density (PSD), which refers to the power at any given frequency, and is often computed from a discrete Fourier transform (DFT). It is common to estimate a smoother PSD from an EEG signal using the Welch method. This method begins with a moving window technique (segments the time series data), wherein FFT is carried out to calculate a modified periodogram of each segment, and then averages the resulting PSD estimates \cite{welch1967use}. Mathematically, Welch power spectral density estimation is:

\begin{equation}
P_w(f)=\frac{1}{L} \sum_{i=0}^{L-1} S_{x x}^i(f)
\end{equation}
\begin{equation}
S_{x x}^i(f)=\frac{Y_s}{K \cdot M}\left|\sum_{n=1}^{M-1} x_i(n) \omega(n) e^{-j 2 \pi f n}\right| \wedge^2
\end{equation}
where  \( S_{x x}^i(f) \) is the optimized periodogram,\( (f = fs)\) is the normalized frequency variable, \((Ys)\) is the scaling factor, \( \omega(n) \) is windowing function, \(L\) is the length of the mark and \( K \) is the normalized constant defined as.

\begin{equation}
\mathrm{K}=\frac{1}{\mathrm{M}} \sum_{\mathrm{n}=0}^{\mathrm{M}-1} \omega^2(\mathrm{n})
\end{equation}
Welch’s method for estimating PSD is influenced by three factors: window length, percentage of overlapping windows, and the number of FFT points. We can also choose different window functions; however, the hanning window is most often preferred due to its good frequency resolution and minimal spectral leakage. We used the hanning window function for calculating the PSD of each of the EEG channels. The average amplitude and band power of the alpha, beta, delta, theta, and gamma waves were also computed.

In the extraction of time-frequency characteristics, the discrete wavelet transform (DWT) is used. The DWT, undoubtedly, is one of the most comprehensive methods for analyzing non-stationary EEG signals \cite{alturki2020eeg}. It approximates the signal as an infinite series of wavelets derived from the mother wavelet making the signal a linear combination of wavelet functions and weighted coefficients of the mother wavelet \(\varphi (t)\). The equation that defines DWT decomposition is as follows:

\begin{equation}
f(t)=\sum_{k=+\infty}^{k=-\infty} C_{n, k} \emptyset\left(2^{-n} t-k\right)+\sum_{k=-\infty}^{k=+\infty} 2^{-\frac{j}{2}} d_{j, k} \varphi\left(2^{-j} t-k\right)
\end{equation}

where \( d_(j,k)\) and \(C_(n,k)\)  represent the approximation and detail coefficients, respectively, \(n\) is the level and \(\varphi\) is the function of scale. 

The first approximation is decomposed, and the process is repeated. At the end of the process, the number of decomposed signals is n+1. For spectral analysis of the EEG data from 16 channels, we employed the Daubechies 4 (db4) as the mother wavelet function, which was applied to decompose the data up to level 4 because it provides the best characteristics for signal features that are classified successfully. For each channel, the detail coefficients (cD1 to cD8) and the approximation coefficients (cA8) were computed, and the wavelet decomposition signal was reconstructed using these characteristics. 

\subsubsection{Case Preparation}
\justifying
A total of 416 features were extracted for each of the 13 tasks i.e. Baseline, Imagery movements (CLH, CRH, PRF, PLF, DRF, DFL), and motor movements (CLH, CRH, PRF, PLF, DRF, DFL). Overall, there were 160 [10x16] time domain features, 96 [1x16 +5x16] pure frequency domain features and [5x16 + 5x16] 160 Wavelet based features.  These features along with output were saved in a CSV file. When it comes to performing binary classification, we sorted the data into eleven cases.
•	Case 1 - All motor activities CLH, CRH, DLF, PLF, DRF, and PRF were classified into one group against baseline Both Eyes Open (BEO).
•	Case 2 - All upper extremity motor activities CLH or CRH vs. BEO.
•	Case 3 - All lower extremity motor activities DLF, PLF, DRF, PRF against BEO.
•	Case 4 - Right upper extremity motor activities (CRH) vs. the BEO.
•	Case 5 - Motor/Motor Imagery activities of left upper extremity (CLH) vs. BEO.
•	Case 6 - Right lower extremity motor activities DRF and PRF against BEO.
•	Case 7 - Right lower extremity motor activity (DRF) vs. baseline BEO.
•	Case 8 - Right lower extremity motor activity (PRF) vs. BEO.
•	Case 9 - Left lower extremity motor activities (DLF, PLF) against BEO.
•	Case 10 - Left lower extremity motor activity (DLF) vs. baseline BEO.
•	Case 11 - The left lower extremity motor activity (PLF) vs. baseline BEO.
A total of 11 x 2 = 22 CSV files were prepared, and each row of the CSV file contains the 416 features and the label that indicates the motor/motor imagery task.

\subsubsection{Features Selection}

When working on analyzing EEG signals several difficulties arise: a noisy environment, artifacts arising from multiple sources, and unwanted signal overlapping as a result of multitasking in the brain. Besides, the signal-to-noise ratio in EEG signals is low and the curse of dimensionality \cite{anuragi2024mitigating} can result in poor classification. To overcome this, a few important features for better classification should be selected. In this study, feature reduction using Principal Component Analysis (PCA) \cite{guerrero2021principal} was attempted, three new features that covered 95\% of the variance of previous features were returned by the PCA algorithm but still, the data points of each of the classes were not so well separated leading to poor classification. Hence, the  Maximum Relevance Minimum Redundancy (MRMR) feature selection technique was employed. The mRMR technique selects the most informative features from a dataset while minimizing redundancy among them. The selection completed by the MRMR algorithm does not allow two selected features to be similar and hence incorporates only useful features. The criterion of feature measurement for MRMR can be written mathematically as:
\begin{equation}
M R M R=\max _s(D(S, c)-R(S))
\end{equation}
The equation shows optimal features can be selected with maximal relevance \(D(S, c)\) and minimal redundancy \(R(S)\)using mRMR. 
\begin{equation}
D(S, c)=\frac{1}{|S|} \sum_{S_i \in S} I\left(f_i ; c\right)
\end{equation}
Here, \( |S|\) is the number of features in sample feature set \(S = {f1, f2, …fn}\) and \(I ( fi;c )\) is the mutual information between the feature \(fi\) and the class \(c\).
 The mutual information \(I(x;y)\) is defined mathematically as:
\begin{equation}
I(x ; y)=\iint p(x, y) \log \frac{p(x, y)}{p(x) p(y)} d x d y
\end{equation}

where \(Sp(x, y)\) is the joint probability density of random variables x and y, and \(Sp(x)\) and \(Sp(y)\) are the marginal probability densities of x and y, respectively. The redundancy of all features in S is the mean of all mutual information between feature \(fi\) and feature \(fj\).
\begin{equation}
R(S)=\frac{1}{|S|^2} \sum_{f_i, f_j \in S} I\left(f_i ; f_j\right)
\end{equation}
MRMR was to select the best four features and this helped in reducing the dimensionality of the data in the dataset to a great extent. The selected features were both informative and diverse as indicated in Figure 5. It is a 3-D plot for case I. Features i.e. 73 (Shannon Entropy of signals acquired from Channel 7), 302 (a wavelet-based feature of the signals acquired from Channel 14) and 358 (a wavelet-based feature of signals acquired from Channel 6) were the top best returned by MRMR algorithm. Yellow colored data points indicate motor activity while blue colored data points represent BEO.
\begin{figure}[h]
    \centering
    \includegraphics[width=0.8\textwidth]{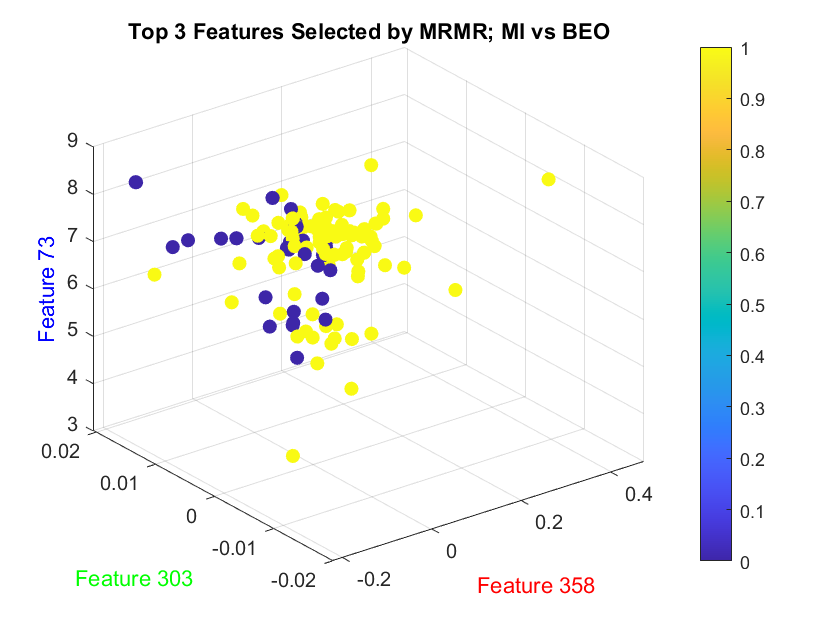} 
    \captionsetup{font=scriptsize} 
    \caption{Best Features selected by MRMR Technique, 0 represents BEO, 1 indicates Motor Activity}
\end{figure}

\subsection{Training and Evaluation of the Machine Learning Algorithms}
The selected features were used to train the Support Vector Machine with Gaussian and Polynomial Kernel, K Nearest Neighbors (with both Euclidean and Mahalanobis distance),  Naïve Bayes and Decision Tree Classifier.
The (KNN) classifier is a simple, non-parametric, supervised learning algorithm widely used for classification problems. It identifies the ‘k’ closest data points (neighbours) to a given test data point in the feature space using distance metrics such as Euclidean Distance, Mahalanobis Distance, etc. It classifies the data point in that class to which the most neighbours belong.  The KNN model does not have training complexity, it calculates distance only when the test data point is provided. In most cases, we choose k (number of neighbours) through Equation 9:

\begin{equation}
k=\sqrt{\left(\frac{N}{2}\right)}
\end{equation}
Where N is the total number of samples \cite{jawed2019classification}.

  The other classifier employed in this work is the Support Vector Machine (SVM). This is probably best suited for binary classification. It classifies data efficiently because it identifies the largest possible margin of the hyperplane which is used to place all the points of a particular class far from any point in other classes. The Naïve Bayes classifier is a machine learning algorithm based on Bayes' theorem that assumes feature independence. The Bayes Theorem gives us information on changes in probability given a particular situation. This algorithm works in linear time and is very scalable and adaptable \cite{cichosz2015naive}. A Decision Tree may be described as “a construct which receives an object or situation characterized in terms of a set of variables and makes a yes/no decision” \cite{russell2016artificial}. The whole decision tree is built based on a recursive splitting of given data into purer subsets according to a set of tests applied to one or many attribute values at each split or node of the tree. The best fact about the decision tree classifiers is that they can perform the feature selection and reduction of the model complexity automatically and the tree architecture gives interpretable information about the strength of the predictive or generation ability of the classification. Models are generally evaluated using hold-out and K fold cross-validation techniques; in hold-out, the validation data is portioned into training (usually 70\%), validating (usually 20\%), and testing set (usually 10\%). The model is trained on the training set and its performance is evaluated by using the validating set while the testing set is used for better generalization purposes. In the K fold cross-validation technique, the dataset is divided into (k) equally sized subsets (folds). The model is trained using (k-1) folds for training and the remaining fold for testing. This process is repeated (k) times, with each fold used exactly once as the test set. As our final dataset contains <250 data points, we used a 10-fold cross-validation technique for evaluating trained models because, on small datasets, K-fold cross-validation is a better technique than hold-out validation \cite{hawkins2003assessing}. The performance of the model was finally reported in the form of average training accuracy, average testing accuracy, precision, recall and F1 Score.

\FloatBarrier
\section{Results}
\FloatBarrier
Tables 2 and 3 show that During Motor tasks, the overall highest Performance with a testing accuracy of 92.5\% and an F1 score of 85.18\% was obtained using our trained SVM Model with Gaussian Kernel as depicted in Figure 6. While during the Motor Imagery tasks, we obtained the highest testing accuracy of 95\% and an F1 score of 87.04\% by using the same SVM Model with a Gaussian Kernel illustrated in Figure 7.

\begin{table}[H]
    \centering
    \caption{Classification results for motor tasks vs BEO}
\resizebox{\textwidth}{!}{%
\begin{tabular}{llllllll}
\begin{tabular}[c]{@{}l@{}}Cases\\    \\ Motor\end{tabular} & ML Model                & \begin{tabular}[c]{@{}l@{}}Training Accuracy\\    \\ (\%)\end{tabular} & \begin{tabular}[c]{@{}l@{}}Testing Accuracy\\    \\ (\%)\end{tabular} & \begin{tabular}[c]{@{}l@{}}Precision\\    \\ (\%)\end{tabular} & \begin{tabular}[c]{@{}l@{}}Recall\\    \\ (\%)\end{tabular} & \begin{tabular}[c]{@{}l@{}}F1 Score\\    \\ (\%)\end{tabular} & Previous Work Highest Accuracy                     \\
\multirow{4}{*}{Case I}                                     & SVM   Gaussian Kernel   & 94.03                                                                  & 92.50                                                                 & 85.17                                                          & 90.80                                                       & 85.18                                                         & \multirow{4}{*}{Less   than (40.0 \%)}             \\
                                                            & KNN   Mahalanobis (N:5) & 93.61                                                                  & 90                                                                    & 73.87                                                          & 83.31                                                       & 76.01                                                         &                                                    \\
                                                            & Naïve   Bayes           & 92.22                                                                  & 87.50                                                                 & 69.95                                                          & 71.58                                                       & 69.58                                                         &                                                    \\
                                                            & Tree   (OVA by Class    & 95.56                                                                  & 88.75                                                                 & 74.63                                                          & 78.61                                                       & 75.40                                                         &                                                    \\
\multirow{4}{*}{Case II}                                    & SVM   Polynomial Kernel & 93.33                                                                  & 86.25                                                                 & 83.00                                                          & 83.54                                                       & 82.16                                                         & \multirow{4}{*}{Ensemble   Subspace KNN (54.7 \%)} \\
                                                            & KNN   Mahalanobis (N:4) & 84.72                                                                  & 81.25                                                                 & 77.75                                                          & 78.74                                                       & 77.58                                                         &                                                    \\
                                                            & Naïve   Bayes           & 92.36                                                                  & 81.25                                                                 & 77.08                                                          & 80.52                                                       & 77.20                                                         &                                                    \\
                                                            & Tree                    & 92.67                                                                  & 72.5                                                                  & 70.08                                                          & 72.32                                                       & 69.03                                                         &                                                    \\
\multirow{4}{*}{Case III}                                   & SVM   Gaussian  Kernel  & 87.58                                                                  & 81.37                                                                 & 74.32                                                          & 76.1                                                        & 72.53                                                         & \multirow{4}{*}{Fine   KNN (43.7 \%)}              \\
                                                            & KNN   Mahalanobis (N:4) & 80.74                                                                  & 79.89                                                                 & 73.35                                                          & 75.12                                                       & 71.13                                                         &                                                    \\
                                                            & Naïve   Bayes           & 86.91                                                                  & 79.99                                                                 & 78.32                                                          & 78,32                                                       & 77.32                                                         &                                                    \\
                                                            & Tree                    & 93.58                                                                  & 82.14                                                                 & 77.98                                                          & 80.57                                                       & 77.57                                                         &                                                    \\
\multirow{4}{*}{Case IV}                                    & SVM                     & 93.89                                                                  & 76.67                                                                 & 78.08                                                          & 73.92                                                       & 70.75                                                         & \multirow{4}{*}{Ensemble Subspace KNN   (73.3 \%)} \\
                                                            & KNN   Mahalanobis (N:9) & 79.29                                                                  & 78.33                                                                 & 73.67                                                          & 70.92                                                       & 70.22                                                         &                                                    \\
                                                            & Naïve   Bayes           & 88.52                                                                  & 83.33                                                                 & 82.58                                                          & 76.25                                                       & 76.79                                                         &                                                    \\
                                                            & Tree                    & 95.38                                                                  & 86.67                                                                 & 86.50                                                          & 87.50                                                       & 84.36                                                         &                                                    \\
\multirow{4}{*}{Case    V}                                  & SVM   Gaussian Kernel   & 97.27                                                                  & 87.86                                                                 & 87.83                                                          & 88.42                                                       & 85.93                                                         & \multirow{4}{*}{Ensemble Subspace KNN   (74.6 \%)} \\
                                                            & KNN   Mahalanobis (N:4) & 87.36                                                                  & 85.00                                                                 & 82.75                                                          & 86.25                                                       & 81.81                                                         &                                                    \\
                                                            & Naïve   Bayes           & 97.95                                                                  & 85.24                                                                 & 83.58                                                          & 81.49                                                       & 80.74                                                         &                                                    \\
                                                            & Tree                    & 92.82                                                                  & 80.48                                                                 & 82                                                             & 80.67                                                       & 77.83                                                         &                                                    \\
\multirow{4}{*}{Case VI}                                    & SVM   Gaussian          & 89.58                                                                  & 77.50                                                                 & 70.17                                                          & 71.28                                                       & 68.45                                                         & \multirow{4}{*}{Fine   KNN (58.3 \%)}              \\
                                                            & KNN   Mahalanobis (N:6) & 78.19                                                                  & 76.25                                                                 & 74.99                                                          & 79.72                                                       & 72.87                                                         &                                                    \\
                                                            & Naïve   Bayes           & 89.07                                                                  & 80.00                                                                 & 72.58                                                          & 74.08                                                       & 70.76                                                         &                                                    \\
                                                            & Tree                    & 90.74                                                                  & 78.33                                                                 & 74.92                                                          & 76.92                                                       & 72.28                                                         &                                                    \\
\multirow{4}{*}{Case VII}                                   & SVM   Polynomial (3rd)  & 97.96                                                                  & 80                                                                    & 77.67                                                          & 76.5                                                        & 73.27                                                         & \multirow{4}{*}{SVM Cubic SVM (73.8 \%)}           \\
                                                            & KNN   Euclidean (N:7)   & 80.93                                                                  & 76.67                                                                 & 68.42                                                          & 70.50                                                       & 65.97                                                         &                                                    \\
                                                            & Naïve   Bayes           & 89.17                                                                  & 76.25                                                                 & 70.57                                                          & 78.38                                                       & 70.71                                                         &                                                    \\
                                                            & Tree                    & 90.97                                                                  & 82.50                                                                 & 78.37                                                          & 80.19                                                       & 77.78                                                         &                                                    \\
\multirow{4}{*}{Case VIII}                                  & SVM   Polynomial (5th)  & 100.00                                                                 & 83.33                                                                 & 82.00                                                          & 80.00                                                       & 78.06                                                         & \multirow{4}{*}{Ensemble Subspace KNN   (73.3 \%)} \\
                                                            & KNN   Mahalanobis (N:5) & 86.67                                                                  & 85.00                                                                 & 79.92                                                          & 80.00                                                       & 77.09                                                         &                                                    \\
                                                            & Naïve   Bayes           & 93.70                                                                  & 76.67                                                                 & 69.25                                                          & 70.25                                                       & 67.46                                                         &                                                    \\
                                                            & Tree                    & 93.89                                                                  & 71.67                                                                 & 70.08                                                          & 73.58                                                       & 66.40                                                         &                                                    \\
\multirow{4}{*}{Case IX}                                    & SVM Gaussian            & 89.67                                                                  & 82.36                                                                 & 82.50                                                          & 83.20                                                       & 81.25                                                         & \multirow{4}{*}{Cubic   SVM (55.6 \%)}             \\
                                                            & KNN   Mahalanobis (N:2) & 89.68                                                                  & 83.61                                                                 & 84.33                                                          & 84.37                                                       & 82.57                                                         &                                                    \\
                                                            & Naïve   Bayes           & 86.80                                                                  & 78.89                                                                 & 79.93                                                          & 81.45                                                       & 78.07                                                         &                                                    \\
                                                            & Tree                    & 90.33                                                                  & 74.03                                                                 & 72.83                                                          & 77.18                                                       & 71.36                                                         &                                                    \\
\multirow{4}{*}{Case X}                                     & SVM   Polynomial (5th)  & 94.70                                                                  & 80.71                                                                 & 78.08                                                          & 77.92                                                       & 74.66                                                         & \multirow{4}{*}{Medium Neural Network   (73.3 \%)} \\
                                                            & KNN   Mahalanobis (N:5) & 82.06                                                                  & 74.05                                                                 & 72.17                                                          & 66.77                                                       & 65.02                                                         &                                                    \\
                                                            & Naïve   Bayes (         & 90.44                                                                  & 75.71                                                                 & 70.50                                                          & 66.33                                                       & 64.09                                                         &                                                    \\
                                                            & Tree                    & 91.67                                                                  & 80.00                                                                 & 77.42                                                          & 78.67                                                       & 73.80                                                         &                                                    \\
\multirow{4}{*}{Case XI}                                    & SVM Gaussian            & 93.68                                                                  & 82.86                                                                 & 81.50                                                          & 81.83                                                       & 76.91                                                         & \multirow{4}{*}{Fine   KNN (67.9 \%)}              \\
                                                            & KNN   Mahalanobis (N:2) & 85.64                                                                  & 84.52                                                                 & 82.75                                                          & 82.33                                                       & 78.72                                                         &                                                    \\
                                                            & Naïve   Bayes           & 91.62                                                                  & 73.33                                                                 & 70.75                                                          & 68.83                                                       & 66.69                                                         &                                                    \\
                                                            & Tree                    & 93.85                                                                  & 78.81                                                                 & 73.42                                                          & 72.58                                                       & 67.61                                                         &                                                   
\end{tabular}%
}

\end{table}
\vspace{7cm} 
\begin{table}[!h]
\caption{Classification results for Imagery tasks vs BEO}
\label{tab:my-table}
\resizebox{\textwidth}{!}{%
\begin{tabular}{llllllll}
\begin{tabular}[c]{@{}l@{}}Cases\\    \\ Motor\end{tabular} & ML Model                & \begin{tabular}[c]{@{}l@{}}Training Accuracy\\    \\ (\%)\end{tabular} & \begin{tabular}[c]{@{}l@{}}Testing Accuracy\\    \\ (\%)\end{tabular} & \begin{tabular}[c]{@{}l@{}}Precision\\    \\ (\%)\end{tabular} & \begin{tabular}[c]{@{}l@{}}Recall\\    \\ (\%)\end{tabular} & \begin{tabular}[c]{@{}l@{}}F1 Score\\    \\ (\%)\end{tabular} & Previous Work Highest Accuracy                     \\
\multirow{4}{*}{Case I}                                     & SVM   Gaussian Kernel   & 94.03                                                                  & 92.50                                                                 & 85.17                                                          & 90.80                                                       & 85.18                                                         & \multirow{4}{*}{Less   than (40.0 \%)}             \\
                                                            & KNN   Mahalanobis (N:5) & 93.61                                                                  & 90                                                                    & 73.87                                                          & 83.31                                                       & 76.01                                                         &                                                    \\
                                                            & Naïve   Bayes           & 92.22                                                                  & 87.50                                                                 & 69.95                                                          & 71.58                                                       & 69.58                                                         &                                                    \\
                                                            & Tree   (OVA by Class    & 95.56                                                                  & 88.75                                                                 & 74.63                                                          & 78.61                                                       & 75.40                                                         &                                                    \\
\multirow{4}{*}{Case II}                                    & SVM   Polynomial Kernel & 93.33                                                                  & 86.25                                                                 & 83.00                                                          & 83.54                                                       & 82.16                                                         & \multirow{4}{*}{Ensemble   Subspace KNN (54.7 \%)} \\
                                                            & KNN   Mahalanobis (N:4) & 84.72                                                                  & 81.25                                                                 & 77.75                                                          & 78.74                                                       & 77.58                                                         &                                                    \\
                                                            & Naïve   Bayes           & 92.36                                                                  & 81.25                                                                 & 77.08                                                          & 80.52                                                       & 77.20                                                         &                                                    \\
                                                            & Tree                    & 92.67                                                                  & 72.5                                                                  & 70.08                                                          & 72.32                                                       & 69.03                                                         &                                                    \\
\multirow{4}{*}{Case III}                                   & SVM   Gaussian  Kernel  & 87.58                                                                  & 81.37                                                                 & 74.32                                                          & 76.1                                                        & 72.53                                                         & \multirow{4}{*}{Fine   KNN (43.7 \%)}              \\
                                                            & KNN   Mahalanobis (N:4) & 80.74                                                                  & 79.89                                                                 & 73.35                                                          & 75.12                                                       & 71.13                                                         &                                                    \\
                                                            & Naïve   Bayes           & 86.91                                                                  & 79.99                                                                 & 78.32                                                          & 78,32                                                       & 77.32                                                         &                                                    \\
                                                            & Tree                    & 93.58                                                                  & 82.14                                                                 & 77.98                                                          & 80.57                                                       & 77.57                                                         &                                                    \\
\multirow{4}{*}{Case IV}                                    & SVM                     & 93.89                                                                  & 76.67                                                                 & 78.08                                                          & 73.92                                                       & 70.75                                                         & \multirow{4}{*}{Ensemble Subspace KNN   (73.3 \%)} \\
                                                            & KNN   Mahalanobis (N:9) & 79.29                                                                  & 78.33                                                                 & 73.67                                                          & 70.92                                                       & 70.22                                                         &                                                    \\
                                                            & Naïve   Bayes           & 88.52                                                                  & 83.33                                                                 & 82.58                                                          & 76.25                                                       & 76.79                                                         &                                                    \\
                                                            & Tree                    & 95.38                                                                  & 86.67                                                                 & 86.50                                                          & 87.50                                                       & 84.36                                                         &                                                    \\
\multirow{4}{*}{Case    V}                                  & SVM   Gaussian Kernel   & 97.27                                                                  & 87.86                                                                 & 87.83                                                          & 88.42                                                       & 85.93                                                         & \multirow{4}{*}{Ensemble Subspace KNN   (74.6 \%)} \\
                                                            & KNN   Mahalanobis (N:4) & 87.36                                                                  & 85.00                                                                 & 82.75                                                          & 86.25                                                       & 81.81                                                         &                                                    \\
                                                            & Naïve   Bayes           & 97.95                                                                  & 85.24                                                                 & 83.58                                                          & 81.49                                                       & 80.74                                                         &                                                    \\
                                                            & Tree                    & 92.82                                                                  & 80.48                                                                 & 82                                                             & 80.67                                                       & 77.83                                                         &                                                    \\
\multirow{4}{*}{Case VI}                                    & SVM   Gaussian          & 89.58                                                                  & 77.50                                                                 & 70.17                                                          & 71.28                                                       & 68.45                                                         & \multirow{4}{*}{Fine   KNN (58.3 \%)}              \\
                                                            & KNN   Mahalanobis (N:6) & 78.19                                                                  & 76.25                                                                 & 74.99                                                          & 79.72                                                       & 72.87                                                         &                                                    \\
                                                            & Naïve   Bayes           & 89.07                                                                  & 80.00                                                                 & 72.58                                                          & 74.08                                                       & 70.76                                                         &                                                    \\
                                                            & Tree                    & 90.74                                                                  & 78.33                                                                 & 74.92                                                          & 76.92                                                       & 72.28                                                         &                                                    \\
\multirow{4}{*}{Case VII}                                   & SVM   Polynomial (3rd)  & 97.96                                                                  & 80                                                                    & 77.67                                                          & 76.5                                                        & 73.27                                                         & \multirow{4}{*}{SVM Cubic SVM (73.8 \%)}           \\
                                                            & KNN   Euclidean (N:7)   & 80.93                                                                  & 76.67                                                                 & 68.42                                                          & 70.50                                                       & 65.97                                                         &                                                    \\
                                                            & Naïve   Bayes           & 89.17                                                                  & 76.25                                                                 & 70.57                                                          & 78.38                                                       & 70.71                                                         &                                                    \\
                                                            & Tree                    & 90.97                                                                  & 82.50                                                                 & 78.37                                                          & 80.19                                                       & 77.78                                                         &                                                    \\
\multirow{4}{*}{Case VIII}                                  & SVM   Polynomial (5th)  & 100.00                                                                 & 83.33                                                                 & 82.00                                                          & 80.00                                                       & 78.06                                                         & \multirow{4}{*}{Ensemble Subspace KNN   (73.3 \%)} \\
                                                            & KNN   Mahalanobis (N:5) & 86.67                                                                  & 85.00                                                                 & 79.92                                                          & 80.00                                                       & 77.09                                                         &                                                    \\
                                                            & Naïve   Bayes           & 93.70                                                                  & 76.67                                                                 & 69.25                                                          & 70.25                                                       & 67.46                                                         &                                                    \\
                                                            & Tree                    & 93.89                                                                  & 71.67                                                                 & 70.08                                                          & 73.58                                                       & 66.40                                                         &                                                    \\
\multirow{4}{*}{Case IX}                                    & SVM Gaussian            & 89.67                                                                  & 82.36                                                                 & 82.50                                                          & 83.20                                                       & 81.25                                                         & \multirow{4}{*}{Cubic   SVM (55.6 \%)}             \\
                                                            & KNN   Mahalanobis (N:2) & 89.68                                                                  & 83.61                                                                 & 84.33                                                          & 84.37                                                       & 82.57                                                         &                                                    \\
                                                            & Naïve   Bayes           & 86.80                                                                  & 78.89                                                                 & 79.93                                                          & 81.45                                                       & 78.07                                                         &                                                    \\
                                                            & Tree                    & 90.33                                                                  & 74.03                                                                 & 72.83                                                          & 77.18                                                       & 71.36                                                         &                                                    \\
\multirow{4}{*}{Case X}                                     & SVM   Polynomial (5th)  & 94.70                                                                  & 80.71                                                                 & 78.08                                                          & 77.92                                                       & 74.66                                                         & \multirow{4}{*}{Medium Neural Network   (73.3 \%)} \\
                                                            & KNN   Mahalanobis (N:5) & 82.06                                                                  & 74.05                                                                 & 72.17                                                          & 66.77                                                       & 65.02                                                         &                                                    \\
                                                            & Naïve   Bayes (         & 90.44                                                                  & 75.71                                                                 & 70.50                                                          & 66.33                                                       & 64.09                                                         &                                                    \\
                                                            & Tree                    & 91.67                                                                  & 80.00                                                                 & 77.42                                                          & 78.67                                                       & 73.80                                                         &                                                    \\
\multirow{4}{*}{Case XI}                                    & SVM Gaussian            & 93.68                                                                  & 82.86                                                                 & 81.50                                                          & 81.83                                                       & 76.91                                                         & \multirow{4}{*}{Fine   KNN (67.9 \%)}              \\
                                                            & KNN   Mahalanobis (N:2) & 85.64                                                                  & 84.52                                                                 & 82.75                                                          & 82.33                                                       & 78.72                                                         &                                                    \\
                                                            & Naïve   Bayes           & 91.62                                                                  & 73.33                                                                 & 70.75                                                          & 68.83                                                       & 66.69                                                         &                                                    \\
                                                            & Tree                    & 93.85                                                                  & 78.81                                                                 & 73.42                                                          & 72.58                                                       & 67.61                                                         &                                                   
\end{tabular}%
}
\end{table}


\begin{figure}[H]
    \centering
    \begin{subfigure}[b]{0.45\textwidth}
        \centering
        \includegraphics[width=\textwidth]{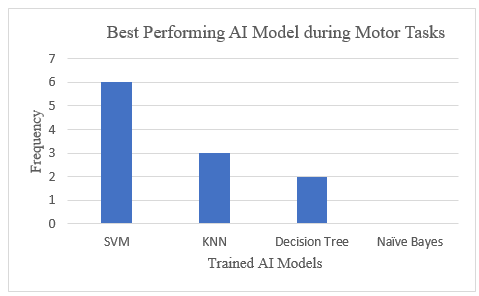} 
        \captionsetup{font=scriptsize} 
        \caption{Best Performing AI Model during Motor Tasks}
    \end{subfigure}
    \hfill
    \begin{subfigure}[b]{0.45\textwidth}
        \centering
        \includegraphics[width=\textwidth]{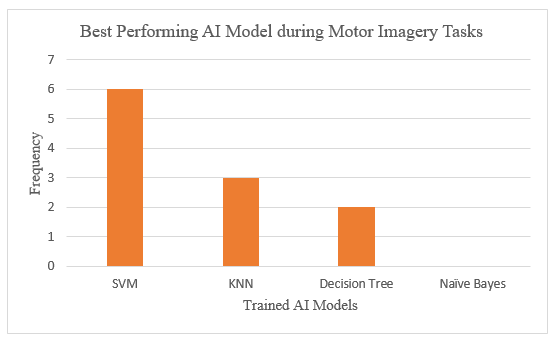} 
        \captionsetup{font=scriptsize} 
        \caption{Best Performing AI Model during Motor Imagery Tasks}
    \end{subfigure}
\end{figure}

\vspace{4cm}
\section{Discussion}
The main goal of this research work was to enhance the accuracy of ML models trained on the MILimbEEG: dataset and also to carry out an in-depth analysis of the ML models’ performance on this dataset for better understanding and classification. Our hypothesis was that Feature Engineering techniques and better optimization can prove valuable for increasing the performance of Machine Learning Models. Publicly available 16 channels of EEG data with labels were used and after pre-processing the time domain, frequency domain, and time-frequency domain features were extracted. The Maximum Relevance Minimum Redundancy  (mRMR) Algorithm was used for selecting the top four features and four AI models were trained on these features and given labels. These features are based on those EEG signals that were acquired through the electrodes located in the sensorimotor cortex, central sulcus, and frontal lobe thus the signals obtained through these parts were more reliable for MA vs BEO classification. After training, models were evaluated using reliable performance metrics such as training accuracy, testing accuracy, precision, recall, and F1 Score.  

  The performance of all the classifier models. was satisfactory and supported our hypotheses. The testing accuracy of the SVM model with Gaussian Kernel was the highest at 92.5\% during motor activities and 95.48\% while performing motor imagery tasks, which means it can generalize to unseen data and classify new signals efficiently. For unbalanced classes, the confusion matrix can better indicate Type I (classifier predicted relevant Brain Activity but it was not) and Type II errors (classifier predicted no brain activity but the person performed it) made by the model. Precision and recall both indicate accuracy, but a significant difference between them exists. Precision is defined as the percentage of the relevant results, whereas recall refers to the percentage of the total of correctly classified relevant results by the algorithm. Recall is useful when the cost of false negatives (FN) is high, while precision is significant when the cost of false positives (FP) is high. As it is impossible to maximize both metrics simultaneously, the F1 score (which is the harmonic mean of both precision and recall) was also used because it integrates precision and recall into a single metric to gain a better understanding of model performance. With a recall score of 88.34\%, a precision score of 86.74\% and an F1 score of 87.04\%, the SVM model proved very reliable in classifying both motor and motor imagery tasks as it will make very few Type I and Type II errors. During Imagery task classifications, more than 80\% testing accuracy was obtained in all cases using only four features out of 416. It was also observed that ML models faced difficulty in classifying imagery dorsiflexion of the right foot from baseline. It can be due to the low signal-to-noise ratio of motor imagery-based EEG signals. K Nearest Neighbors algorithm also performed very well in both types of activities. As the EEG dataset is non-linear, the distance metric such as the euclidean distance generally does not give good results as it is more suitable for linear datasets and researchers mostly prefer the Mahalanobis distance formula having a covariance parameter. But after Feature reduction, it can be seen that both types of distance metrics can ensure better classification of EEG signals for BCI-based applications. The performance of Naïve Bayes and Decision Tree was also satisfactory and can be further enhanced by fine-tuning.

\cite{handiru2016optimized} attempted to make the motor imagery (MI) classification less computationally complex by reducing the number of EEG channels.  The authors introduce a method known as Iterative Multi-objective Optimization for Channel Selection (IMOCS) for choosing the most useful EEG channels for motor imagery classification tasks. They achieved an average classification accuracy of 63 per cent for 85 subjects. The trained models in this work achieved far better accuracy through the above-discussed Feature engineering techniques.

\cite{xu2014classification} utilized the combination of two features (Autoregressive and local Binary Pattern) with a Gradient Boosting algorithm in combination with ordinary least squares (OLS) regression for MI classification of ECoG based BCI system. They selected 14/64 channels whose final cross-validation accuracies were higher than 66\% and obtained the best accuracy of 93\%. However, ECoG-based BCI systems are invasive and in this work by utilizing the mRMR algorithm, it was proved that the same accuracy can be obtained in a non-invasive way by using EEG signals.

For channel selection, the sequential floating forward selection (SFFS) was designed by \cite{pudil1994floating}. However, it was very slow especially if the number of features is very large. To solve this problem,  \cite{qiu2016improved} proposed Improved Simulated Fast Forward Selection (SFFS) algorithms to select the optimal channels. It could select or delete several channels at each time. To evaluate the performance of the improved SFFS algorithm, two BCI datasets from publicly available BCI competitions are considered. In the study, Dataset I, which has 59 channels, and Dataset II, with 118 channels were used. They assessed the model by iteratively removing channels and achieved a mean accuracy of 78\% using Dataset I and 83 \% using Dataset II. In the present study, four relevant features from four channels were chosen to reduce complexity, and promising and reliable results were obtained. Moreover, this study offers richer results for assessing the practicability of machine learning models for the given EEG dataset. 
 
\cite{mohammadi2022electroencephalography} also used BCI Competition IV - datasets 2a. In their work, Common Spatial Patterns (CSP) for dimensional reduction were used, and temporal and time-frequency features were extracted. They obtained a mean classification accuracy of 64\%. They trained the model based on EEG data of the same subject instead of different subjects and thus it will not yield better generalization for classifying unknown real-world EEG data and hence it cannot prove significant for online BCI systems. Our model is trained on the dataset that was obtained from 24 individuals in which subjects suffering from neurological disorders were also there and through performance evaluation, these models will lead to better generalization and this work also shows the importance of the MI Limb EEG dataset for designing of the online BCI Systems.

  Recently, \cite{olvera2024eeg} showed that for selecting features in EEG-based motor imagery classification, the practical feasibility of a novel genetic quantum algorithm (NQGA) is suitable. They employed the BCI Competition IV dataset 2b and adopted the wrapper feature selection method with a quantum genetic algorithm. They also used a combination of a classical and quantum network with a baseline feature extraction network.  In the classification process, a Variational Quantum Circuit was used and the mean accuracy of 83.82\%, was obtained when using the subject-dependent cross-classification approach. Although it is a very impressive study and opened new doors of research, we cannot rely only on a subject-dependent cross-classification approach for designing the online BCI systems. 
  
So, a comparison with the above results shows that by using the MI Limb EEG dataset, and novel feature selection approaches such as NQGA, better BCI systems can be developed for online Motor Imagery Classification and can be used for Neuro-rehabilitation purposes.

\section{Conclusion}
The main goal of this research work was to improve the accuracy of Machine Learning Models for classifying Motor/Motor Imagery tasks and rest so that better, reliable, less complex and cost-effective BCI systems can be developed for neuro-rehabilitation. It was proved through this work that the MI Limb EEG dataset along with efficient feature engineering methods can improve the accuracy and F1 Score of Machine Learning Models for MI BCI applications. The model evaluation provided a thorough understanding of the performance and we can conclude from it that SVM is a powerful model to deal with the non-linearity in EEG datasets. KNN Model can also be used for designing online BCI systems by implementing reliable feature selection techniques and the  Maximum Relevance Minimum Redundancy(MRMR) feature reduction algorithm. Our final model provides better generalization because it is trained on a comparatively more diverse dataset and in future, we will integrate it with hardware for controlling neuro-prostheses to help patients suffering from motor disabilities.

\subsection*{CRedit authorship contribution statement}
Syed Saim Gardezi: Conceptualisation, Writing – original draft, Methodology, Investigation, Formal analysis. Soyiba Jawed: Methodology, Formal analysis, conceptualisation. Review and editing.  Mahnoor Khan: Writing – original draft, Methodology, Investigation. Muneeba Bukhari: Writing – original draft, Rizwan Ahmed Khan: Formal analysis, Review and editing.
\subsection*{Declaration of competing interest}
The authors declare that they have no known competing financial interests or personal relationships that could have appeared to influence the work reported in this paper.
\subsection*{Appendix A. Supplementary data}
Data used in this research work is available at: MILimbEEG: An EEG Signals Dataset based on Upper and Lower Limb Task During the Execution of Motor and Motorimagery Tasks - Mendeley Data

\bibliographystyle{apalike}

\begin{thebibliography}{}
	
	\bibitem[Al-Saegh et~al., 2021]{ALSAEGH2021102172}
	Al-Saegh, A., Dawwd, S.~A., and Abdul-Jabbar, J.~M. (2021).
	\newblock Deep learning for motor imagery eeg-based classification: A review.
	\newblock {\em Biomedical Signal Processing and Control}, 63:102172.
	
	\bibitem[Alturki et~al., 2020]{alturki2020eeg}
	Alturki, F.~A., AlSharabi, K., Abdurraqeeb, A.~M., and Aljalal, M. (2020).
	\newblock Eeg signal analysis for diagnosing neurological disorders using
	discrete wavelet transform and intelligent techniques.
	\newblock {\em Sensors}, 20(9):2505.
	
	\bibitem[Alzahab et~al., 2021]{alzahab2021hybrid}
	Alzahab, N.~A., Apollonio, L., Di~Iorio, A., Alshalak, M., Iarlori, S.,
	Ferracuti, F., Monteriu, A., and Porcaro, C. (2021).
	\newblock Hybrid deep learning (hdl)-based brain-computer interface (bci)
	systems: a systematic review.
	\newblock {\em Brain sciences}, 11(1):75.
	
	\bibitem[Anuragi et~al., 2024]{anuragi2024mitigating}
	Anuragi, A., Sisodia, D.~S., and Pachori, R.~B. (2024).
	\newblock Mitigating the curse of dimensionality using feature projection
	techniques on electroencephalography datasets: an empirical review.
	\newblock {\em Artificial Intelligence Review}, 57(3):75.
	
	\bibitem[Asanza et~al., 2023]{asanza2023milimbeeg}
	Asanza, V., Lorente-Leyva, L.~L., Peluffo-Ord{\'o}{\~n}ez, D.~H., Montoya, D.,
	and Gonzalez, K. (2023).
	\newblock Milimbeeg: A dataset of eeg signals related to upper and lower limb
	execution of motor and motor imagery tasks.
	\newblock {\em Data in Brief}, 50:109540.
	
	\bibitem[Cichosz, 2015]{cichosz2015naive}
	Cichosz, P. (2015).
	\newblock Na{\"\i}ve bayes classifier.
	
	\bibitem[Cooper et~al., 2006]{cooper2006introduction}
	Cooper, R.~A., Ohnabe, H., and Hobson, D.~A. (2006).
	\newblock {\em An introduction to rehabilitation engineering}.
	\newblock CRC Press.
	
	\bibitem[Donchin et~al., 2000]{donchin2000mental}
	Donchin, E., Spencer, K.~M., and Wijesinghe, R. (2000).
	\newblock The mental prosthesis: assessing the speed of a p300-based
	brain-computer interface.
	\newblock {\em IEEE transactions on rehabilitation engineering}, 8(2):174--179.
	
	\bibitem[Ertel, 2024]{ertel2024introduction}
	Ertel, W. (2024).
	\newblock {\em Introduction to artificial intelligence}.
	\newblock Springer Nature.
	
	\bibitem[Grozea et~al., 2011]{grozea2011bristle}
	Grozea, C., Voinescu, C.~D., and Fazli, S. (2011).
	\newblock Bristle-sensors—low-cost flexible passive dry eeg electrodes for
	neurofeedback and bci applications.
	\newblock {\em Journal of neural engineering}, 8(2):025008.
	
	\bibitem[Guerrero et~al., 2021]{guerrero2021principal}
	Guerrero, M.~C., Parada, J.~S., and Espitia, H.~E. (2021).
	\newblock Principal components analysis of eeg signals for epileptic patient
	identification.
	\newblock {\em Computation}, 9(12):133.
	
	\bibitem[Gwon et~al., 2023]{gwon2023review}
	Gwon, D., Won, K., Song, M., Nam, C.~S., Jun, S.~C., and Ahn, M. (2023).
	\newblock Review of public motor imagery and execution datasets in
	brain-computer interfaces.
	\newblock {\em Frontiers in human neuroscience}, 17:1134869.
	
	\bibitem[Handiru and Prasad, 2016]{handiru2016optimized}
	Handiru, V.~S. and Prasad, V.~A. (2016).
	\newblock Optimized bi-objective eeg channel selection and cross-subject
	generalization with brain--computer interfaces.
	\newblock {\em IEEE Transactions on Human-Machine Systems}, 46(6):777--786.
	
	\bibitem[Hawkins et~al., 2003]{hawkins2003assessing}
	Hawkins, D.~M., Basak, S.~C., and Mills, D. (2003).
	\newblock Assessing model fit by cross-validation.
	\newblock {\em Journal of chemical information and computer sciences},
	43(2):579--586.
	
	\bibitem[Jawed et~al., 2019]{jawed2019classification}
	Jawed, S., Amin, H.~U., Malik, A.~S., and Faye, I. (2019).
	\newblock Classification of visual and non-visual learners using
	electroencephalographic alpha and gamma activities.
	\newblock {\em Frontiers in behavioral neuroscience}, 13:86.
	
	\bibitem[Kaya et~al., 2018]{kaya2018large}
	Kaya, M., Binli, M.~K., Ozbay, E., Yanar, H., and Mishchenko, Y. (2018).
	\newblock A large electroencephalographic motor imagery dataset for
	electroencephalographic brain computer interfaces.
	\newblock {\em Scientific data}, 5(1):1--16.
	
	\bibitem[Khademi et~al., 2023]{khademi2023review}
	Khademi, Z., Ebrahimi, F., and Kordy, H.~M. (2023).
	\newblock A review of critical challenges in mi-bci: From conventional to deep
	learning methods.
	\newblock {\em Journal of Neuroscience Methods}, 383:109736.
	
	\bibitem[Khan et~al., 2020]{khan2020review}
	Khan, M.~A., Das, R., Iversen, H.~K., and Puthusserypady, S. (2020).
	\newblock Review on motor imagery based bci systems for upper limb post-stroke
	neurorehabilitation: From designing to application.
	\newblock {\em Computers in biology and medicine}, 123:103843.
	
	\bibitem[Kumar et~al., 2017]{Kumar}
	Kumar, S., Sharma, A., and Tsunoda, T. (2017).
	\newblock An improved discriminative filter bank selection approach for motor
	imagery eeg signal classification using mutual information.
	\newblock {\em BMC Bioinformatics}, 18.
	
	\bibitem[Mansouri et~al., 2023]{mansouri2023telemetric}
	Mansouri, M.~T., Ahmed, M.~T., Cassim, T.~Z., Kreuzer, M., Graves, M.~C.,
	Fenzl, T., and Garc{\'\i}a, P.~S. (2023).
	\newblock Telemetric electroencephalography recording in anesthetized mice—a
	novel system using minimally-invasive needle electrodes with a wireless
	openbci™ cyton biosensing board.
	\newblock {\em MethodsX}, 10:102187.
	
	\bibitem[Mohammadi et~al., 2022]{mohammadi2022electroencephalography}
	Mohammadi, E., Daneshmand, P.~G., and Khorzooghi, S. M. S.~M. (2022).
	\newblock Electroencephalography-based brain--computer interface motor imagery
	classification.
	\newblock {\em Journal of Medical Signals \& Sensors}, 12(1):40--47.
	
	\bibitem[Murguialday et~al., 2011]{murguialday2011transition}
	Murguialday, A.~R., Hill, J., Bensch, M., Martens, S., Halder, S., Nijboer, F.,
	Schoelkopf, B., Birbaumer, N., and Gharabaghi, A. (2011).
	\newblock Transition from the locked in to the completely locked-in state: a
	physiological analysis.
	\newblock {\em Clinical Neurophysiology}, 122(5):925--933.
	
	\bibitem[Neuper et~al., 2009]{neuper2009motor}
	Neuper, C., Scherer, R., Wriessnegger, S., and Pfurtscheller, G. (2009).
	\newblock Motor imagery and action observation: modulation of sensorimotor
	brain rhythms during mental control of a brain--computer interface.
	\newblock {\em Clinical neurophysiology}, 120(2):239--247.
	
	\bibitem[Nicolas-Alonso and Gomez-Gil, 2012]{nicolas2012brain}
	Nicolas-Alonso, L.~F. and Gomez-Gil, J. (2012).
	\newblock Brain computer interfaces, a review.
	\newblock {\em sensors}, 12(2):1211--1279.
	
	\bibitem[Olvera et~al., 2024]{olvera2024eeg}
	Olvera, C., Ross, O.~M., and Rubio, Y. (2024).
	\newblock Eeg-based motor imagery classification with quantum algorithms.
	\newblock {\em Expert Systems with Applications}, 247:123354.
	
	\bibitem[Organization et~al., 2022]{world2022global}
	Organization, W.~H. et~al. (2022).
	\newblock {\em Global report on health equity for persons with disabilities}.
	\newblock World Health Organization.
	
	\bibitem[Padfield et~al., 2019]{padfield2019eeg}
	Padfield, N., Zabalza, J., Zhao, H., Masero, V., and Ren, J. (2019).
	\newblock Eeg-based brain-computer interfaces using motor-imagery: Techniques
	and challenges.
	\newblock {\em Sensors}, 19(6):1423.
	
	\bibitem[Pudil et~al., 1994]{pudil1994floating}
	Pudil, P., Novovi{\v{c}}ov{\'a}, J., and Kittler, J. (1994).
	\newblock Floating search methods in feature selection.
	\newblock {\em Pattern recognition letters}, 15(11):1119--1125.
	
	\bibitem[Qiu et~al., 2016]{qiu2016improved}
	Qiu, Z., Jin, J., Lam, H.-K., Zhang, Y., Wang, X., and Cichocki, A. (2016).
	\newblock Improved sffs method for channel selection in motor imagery based
	bci.
	\newblock {\em Neurocomputing}, 207:519--527.
	
	\bibitem[Raj et~al., 2020]{raj2020feature}
	Raj, V., Hazarika, J., and Hazra, R. (2020).
	\newblock Feature selection for attention demanding task induced eeg detection.
	\newblock In {\em 2020 IEEE Applied Signal Processing Conference (ASPCON)},
	pages 11--15. IEEE.
	
	\bibitem[Razavi et~al., 2022]{razavi2022opensync}
	Razavi, M., Janfaza, V., Yamauchi, T., Leontyev, A., Longmire-Monford, S., and
	Orr, J. (2022).
	\newblock Opensync: an open-source platform for synchronizing multiple measures
	in neuroscience experiments.
	\newblock {\em Journal of neuroscience methods}, 369:109458.
	
	\bibitem[Rodr{\'\i}guez-Berm{\'u}dez and Garc{\'\i}a-Laencina,
	2012]{rodriguez2012automatic}
	Rodr{\'\i}guez-Berm{\'u}dez, G. and Garc{\'\i}a-Laencina, P.~J. (2012).
	\newblock Automatic and adaptive classification of electroencephalographic
	signals for brain computer interfaces.
	\newblock {\em Journal of medical systems}, 36:51--63.
	
	\bibitem[Russell and Norvig, 2016]{russell2016artificial}
	Russell, S.~J. and Norvig, P. (2016).
	\newblock {\em Artificial intelligence: a modern approach}.
	\newblock Pearson.
	
	\bibitem[Sabrin, 2022]{sabrin2022biopsychosocial}
	Sabrin, S.~M. (2022).
	\newblock {\em Biopsychosocial impact on person with physical disabilities}.
	\newblock PhD thesis, Bangladesh Health Professions Institute, Faculty of
	Medicine, the University~….
	
	\bibitem[Schalk et~al., 2007]{schalk2007decoding}
	Schalk, G., Kubanek, J., Miller, K.~J., Anderson, N., Leuthardt, E.~C.,
	Ojemann, J.~G., Limbrick, D., Moran, D., Gerhardt, L.~A., and Wolpaw, J.~R.
	(2007).
	\newblock Decoding two-dimensional movement trajectories using
	electrocorticographic signals in humans.
	\newblock {\em Journal of neural engineering}, 4(3):264.
	
	\bibitem[Schalk et~al., 2004]{schalk2004bci2000}
	Schalk, G., McFarland, D.~J., Hinterberger, T., Birbaumer, N., and Wolpaw,
	J.~R. (2004).
	\newblock Bci2000: a general-purpose brain-computer interface (bci) system.
	\newblock {\em IEEE Transactions on biomedical engineering}, 51(6):1034--1043.
	
	\bibitem[Schalk et~al., 2008]{schalk2008two}
	Schalk, G., Miller, K.~J., Anderson, N.~R., Wilson, J.~A., Smyth, M.~D.,
	Ojemann, J.~G., Moran, D.~W., Wolpaw, J.~R., and Leuthardt, E.~C. (2008).
	\newblock Two-dimensional movement control using electrocorticographic signals
	in humans.
	\newblock {\em Journal of neural engineering}, 5(1):75.
	
	\bibitem[Schirrmeister et~al., 2017]{schirrmeister2017deep}
	Schirrmeister, R.~T., Springenberg, J.~T., Fiederer, L. D.~J., Glasstetter, M.,
	Eggensperger, K., Tangermann, M., Hutter, F., Burgard, W., and Ball, T.
	(2017).
	\newblock Deep learning with convolutional neural networks for eeg decoding and
	visualization.
	\newblock {\em Human brain mapping}, 38(11):5391--5420.
	
	\bibitem[Seeck et~al., 2017]{seeck2017standardized}
	Seeck, M., Koessler, L., Bast, T., Leijten, F., Michel, C., Baumgartner, C.,
	He, B., and Beniczky, S. (2017).
	\newblock The standardized eeg electrode array of the ifcn.
	\newblock {\em Clinical neurophysiology}, 128(10):2070--2077.
	
	\bibitem[Strehl, 2009]{strehl2009slow}
	Strehl, U. (2009).
	\newblock Slow cortical potentials neurofeedback.
	\newblock {\em Journal of Neurotherapy}, 13(2):117--126.
	
	\bibitem[Suner et~al., 2005]{suner2005reliability}
	Suner, S., Fellows, M.~R., Vargas-Irwin, C., Nakata, G.~K., and Donoghue, J.~P.
	(2005).
	\newblock Reliability of signals from a chronically implanted, silicon-based
	electrode array in non-human primate primary motor cortex.
	\newblock {\em IEEE transactions on neural systems and rehabilitation
		engineering}, 13(4):524--541.
	
	\bibitem[Tangermann et~al., 2012]{tangermann2012review}
	Tangermann, M., M{\"u}ller, K.-R., Aertsen, A., Birbaumer, N., Braun, C.,
	Brunner, C., Leeb, R., Mehring, C., Miller, K.~J., M{\"u}ller-Putz, G.~R.,
	et~al. (2012).
	\newblock Review of the bci competition iv.
	\newblock {\em Frontiers in neuroscience}, 6:55.
	
	\bibitem[van Steensel~Umcu et~al., 2014]{Umcu2014TitleCT}
	van Steensel~Umcu, M., Umcu, G.~K., Blefari, M.~L., Brunner, C., Blankertz, B.,
	H{\"o}hne, J., Ortner, R., and Reuderink, B. (2014).
	\newblock Title: Contribution to roadmap.
	
	\bibitem[Vansteensel et~al., 2017]{vansteensel2017brain}
	Vansteensel, M., Kristo, G., Aarnoutse, E., and Ramsey, N. (2017).
	\newblock The brain-computer interface researcher’s questionnaire: from
	research to application.
	\newblock {\em Brain-Computer Interfaces}, 4(4):236--247.
	
	\bibitem[Wang et~al., 2008]{wang2008brain}
	Wang, Y., Gao, X., Hong, B., Jia, C., and Gao, S. (2008).
	\newblock Brain-computer interfaces based on visual evoked potentials.
	\newblock {\em IEEE Engineering in medicine and biology magazine},
	27(5):64--71.
	
	\bibitem[Welch, 1967]{welch1967use}
	Welch, P. (1967).
	\newblock The use of fast fourier transform for the estimation of power
	spectra: a method based on time averaging over short, modified periodograms.
	\newblock {\em IEEE Transactions on audio and electroacoustics}, 15(2):70--73.
	
	\bibitem[Wolpaw and Wolpaw, 2012]{wolpaw2012brain}
	Wolpaw, J.~R. and Wolpaw, E.~W. (2012).
	\newblock Brain-computer interfaces: something new under the sun.
	\newblock {\em Brain-computer interfaces: principles and practice}, 14:3--12.
	
	\bibitem[Xu et~al., 2014]{xu2014classification}
	Xu, F., Zhou, W., Zhen, Y., and Yuan, Q. (2014).
	\newblock Classification of motor imagery tasks for electrocorticogram based
	brain-computer interface.
	\newblock {\em Biomedical Engineering Letters}, 4:149--157.
	
\end{thebibliography}


\end{document}